\documentclass[prb,twocolumn,superscriptaddress,showpacs,aps]{revtex4-1}

\usepackage{color}
\usepackage{graphicx}
\usepackage{bm}
\usepackage{amsmath}
\usepackage{natbib}

\begin{document}

\newcommand{\Tc}{T_{\text c}}
\newcommand{\Hcii}{H_{\text c2}}
\newcommand{\fq}{\Phi_0}
\newcommand{\lamn}{\lambda_n}
\newcommand{\Gac}{\Gamma_{ac}}
\newcommand{\GVL}{\Gamma_{\text{VL}}}
\newcommand{\GHcii}{\Gamma_{\Hcii}}
\newcommand{\QVLvec}{\bm{Q}_{\text{VL}}}
\newcommand{\QVL}{Q_{\text{VL}}}
\newcommand{\ISF}{I_{\text{SF}}}
\newcommand{\INSF}{I_{\text{NSF}}}
\newcommand{\hxL}{h_x^{\text{London}}}
\newcommand{\hzL}{h_z^{\text{London}}}
\newcommand{\hxPPE}{h_x^{\text{PPE}}}
\newcommand{\hzPPE}{h_z^{\text{PPE}}}
\newcommand{\fPPE}{f_{\text{PPE}}}
\newcommand{\KFA}{KFe$_2$As$_2$}
\newcommand{\MgB}{MgB$_2$}
\newcommand{\SRO}{Sr$_2$RuO$_4$}
\newcommand{\emf}[1] {\textcolor{black}{#1}}
\newcommand{\fh}[1] {\textcolor{black}{#1}}
\newcommand{\hkf}[1] {\textcolor{black}{#1}}
\newcommand{\sjk}[1] {\textcolor{black}{#1}}

\title{Simultaneous Evidence for Pauli Paramagnetic Effects and Multiband Superconductivity in {\KFA} by Small-Angle Neutron Scattering Studies of the Vortex Lattice}

\author{S.~J.~Kuhn}
\altaffiliation{Current address: Oak Ridge National Laboratory, Oak Ridge, Tennessee, USA}
\affiliation{Department of Physics, University of Notre Dame, Notre Dame, Indiana 46556, USA}

\author{H.~Kawano-Furukawa}
\affiliation{Division of Natural/Applied Science, Graduate School of Humanities and Science, Ochanomizu University, Bunkyo-ku, Tokyo 112-8610, Japan}

\author{E.~Jellyman}
\affiliation{School of Physics and Astronomy, University of Birmingham, Birmingham B15 2TT, United Kingdom}

\author{R. Riyat}
\affiliation{School of Physics and Astronomy, University of Birmingham, Birmingham B15 2TT, United Kingdom}

\author{E.~M. Forgan}
\affiliation{School of Physics and Astronomy, University of Birmingham, Birmingham B15 2TT, United Kingdom}

\author{M.~Ono}
\affiliation{Physics department, Ochanomizu University, Bunkyo-ku, Tokyo 112-8610, Japan}

\author{K.~Kihou}
\affiliation{National Institute of Advanced Industrial Science and Technology (AIST), Tsukuba, Ibaraki 305-8568, Japan}

\author{C.~H.~Lee}
\affiliation{National Institute of Advanced Industrial Science and Technology (AIST), Tsukuba, Ibaraki 305-8568, Japan}

\author{F.~Hardy}
\affiliation{Institute for Solid State Physics (IFP), Karlsruhe Institute of Technology, D-76021 Karlsruhe, Germany}

\author{P.~Adelmann}
\affiliation{Institute for Solid State Physics (IFP), Karlsruhe Institute of Technology, D-76021 Karlsruhe, Germany}

\author{Th.~Wolf}
\affiliation{Institute for Solid State Physics (IFP), Karlsruhe Institute of Technology, D-76021 Karlsruhe, Germany}

\author{C.~Meingast}
\affiliation{Institute for Solid State Physics (IFP), Karlsruhe Institute of Technology, D-76021 Karlsruhe, Germany}

\author{J.~Gavilano}
\affiliation{Laboratory for Neutron Scattering, Paul Scherrer Institute, CH-5232 Villigen, Switzerland}

\author{M.~R.~Eskildsen}
\email{Corresponding author: eskildsen@nd.edu}
\affiliation{Department of Physics, University of Notre Dame, Notre Dame, Indiana 46556, USA}

\date{\today}

\begin{abstract}
We study the intrinsic anisotropy of the superconducting state in {\KFA}, using small-angle neutron scattering to image the vortex lattice as the applied magnetic field is rotated towards the FeAs crystalline planes.
The anisotropy is found to be strongly field dependent, indicating multiband superconductivity.
Furthermore, the high field anisotropy significantly exceeds that of the upper critical field, providing further support for Pauli limiting in {\KFA} for field applied in the basal plane.
The effect of Pauli paramagnetism on the unpaired quasiparticles in the vortex cores is directly evident from the ratio of scattered intensities due to the longitudinal and transverse vortex lattice field modulation. 
\end{abstract}


\maketitle

\section{Introduction}
A comprehensive, detailed understanding of the interplay between superconductivity and magnetism is a longstanding problem of great scientific interest. As a result, materials with a particularly strong coupling between the two phenomena continue to attract attention.
Recently, members of the iron-based superconductors, where the Cooper pairing is theorized to arise
from magnetic interactions, have emerged as very good model systems for the study of such effects.\cite{Hirschfeld:2011jy,Chubukov:2012ct}.
Parent compounds such as BaFe$_2$As$_2$ exhibit long range antiferromagnetic ordering which may be suppressed by doping on either the Ba- or Fe-site, giving rise to superconductivity in a manner reminiscent of the high-temperature cuprates.\cite{Johnston:2010cs,RefWorks:789,Stewart:2011kw}

Among the iron-based superconductors {\KFA} is of particular interest.
Strong multiband features, similar to that of {\MgB}, are observed in small-angle neutron scattering\cite{KawanoFurukawa:2011hw} and thermodynamic measurements.\cite{Hardy:2014te}
Furthermore, {\KFA} exhibits strongly renormalized band effective masses,\cite{Terashima:2010iu,Terashima:2013vv} leading to an enhanced Pauli susceptibility.\cite{Hardy:2013fe}
Together with a large Ginzburg-Landau parameter $\kappa = 87$,\cite{Burger:2013tg} this is a prerequisite for Pauli (or paramagnetic) limiting and the possible existence of a spatially inhomogeneous Fulde-Ferrell-Larkin-Ovchinnikov (FFLO) state.\cite{RefWorks:770,Wiesenmayer:2011it,Burger:2013tg,Zocco:2013gz,Kittaka:2014ia,Hardy:2014te,Takahashi:2014iz}
Importantly, single crystals of this compound can be synthesized in much cleaner form than most other iron-pnictide and chalcogenide superconductors,
indicated by the observation of quantum oscillations\cite{Terashima:2010iu,Terashima:2013vv}
and highly reversible magnetization for fields perpendicular as well as within the crystalline basal plane.\cite{Burger:2013tg}
This greatly expands the range of feasible experimental techniques that can be applied to study the superconducting state in {\KFA}.

Superconducting vortices, introduced by an applied magnetic field, can serve as a sensitive probe of the superconducting state in the host material. For more than half a century small-angle neutron scattering (SANS) has been used to study the vortex lattice (VL) in a wide range of materials,\cite{Cribier:1964wr,Schelten:1974vc,RefWorks:761,Forgan:1990ez,RefWorks:38,Riseman:1998to}
and has provided often unique information about gap nodes and dispersion,\cite{RefWorks:215,KawanoFurukawa:2011hw,Gannon:2015ct,RefWorks:742}
non-local effects,\cite{RefWorks:829,RefWorks:34}
multiband superconductivity,\cite{RefWorks:19}
Pauli paramagnetic effects,\cite{RefWorks:6,RefWorks:5,RefWorks:1,Das:2012fb}
and a direct measure of the intrinsic superconducting anisotropy ($\Gac$).\cite{Christen:1985wo,Gammel:1994vn,Kealey:2001fi,Das:2012fb,RefWorks:12,Rastovski:2013hh,KawanoFurukawa:2013cs}
The latter quantity may be directly measured by the field-angle-dependent distortion of the VL structure ($\GVL$) from a regular hexagonal pattern. In London theory $\Gac$  represents the anisotropy of the penetration depth.\cite{Thiemann:1989uw} In Ginzburg-Landau theory it also represents the anisotropy of the coherence length, which can arise from both superconducting gap and Fermi velocity anisotropy.  However, $\Gac$ is particularly important in materials where the upper critical field is Pauli limited along one or more crystalline directions, because then the $\Hcii$ anisotropy differs from the intrinsic superconducting anisotropy.

Here we report on SANS measurements of the VL in {\KFA} that substantially extend previous studies\cite{KawanoFurukawa:2013cs} to higher applied magnetic fields and, importantly, directions closer to the basal plane.
This allows a more precise determination of $\Gac$, which is found to be strongly field dependent indicating multiband superconductivity in this material. Furthermore, the high field anisotropy exceeds that of the upper critical field, providing further support for Pauli limiting in {\KFA} for field applied in the basal plane.
Finally, we are able determine the contribution to the field modulation in the mixed state due to PPEs by measuring both the non-spin flip and spin flip VL scattered intensity. This represents the first instance where all these effects have been observed simultaneously and in a comprehensive manner by a single measurement technique.

\section{Experimental Details}
The SANS measurements were performed using large mosaics of co-aligned {\KFA} single crystals. A total of 3 experiments were carried out, each using a newly prepared mosaic due to the air sensitivity of the crystals. The initial experiment used crystals grown at the Karlsruhe Institute of Technology (KIT).
The more extensive data presented here were obtained using crystals grown at the National Institute of Advanced Industrial Science and Technology (AIST).
The crystals from the two sources provided qualitatively similar results. 

\begin{figure}
 \includegraphics{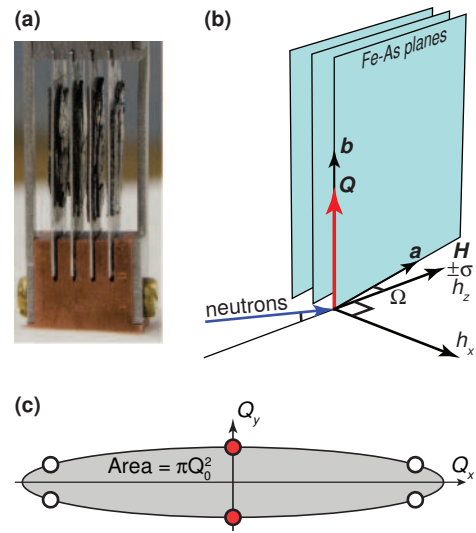}
 \caption{(Color online)
  Sample mosaic (a), experimental geometry (b) and vortex lattice anisotropy (c).
  Panel (a): Photograph showing {\KFA} single crystals mounted on both sides of 4 parallel aluminum plates. The total mass of the crystals in this mosaic is $\sim 2$~g. 
  Panel (b): The coordinate system is defined with $z$ along $\bm{H}$ and $y$ vertical in the Fe-As basal plane (along $\bm{b}$).
  The applied magnetic field ($\bm{H}$) is rotated away from the basal plane by an angle $\Omega$.
  Neutron spins ($\sigma$) are parallel or antiparallel to the magnetic field.
  The incident neutron beam is in the $yz$ plane, at an angle $\varphi$ relative to the field direction.
  The observed VL scattering vector is denoted $\bm{Q}$ and the longitudinal and transverse Fourier component of the field modulation by $h_z$ and $h_x$, respectively.
  Panel (c): Schematic of VL Bragg reflections lying on an ellipse in reciprocal space, with major-to-minor axis ratio given by $\GVL$ (shown here for $\GVL = 6$).
  The area of the ellipse is determined by the applied field, $\pi Q_0^2 = 8\pi^3 \mu_0 H/\sqrt{3}\fq$, and as a result only the filled (red) peaks are required to determine $\GVL$.
  \label{ScatGeom}}
\end{figure}

The crystals were grown in a manner similar to those used in previous SANS experiments,\cite{KawanoFurukawa:2011hw} with critical temperatures $\Tc = 3.4$~K~$\pm \;0.2$~K (10\% - 90\% resistivity range).
The crystals were co-aligned and mounted on parallel aluminum plates to maximize the sample volume while minimizing the sample thickness traversed by the neutrons when the beam direction was close to the basal plane, Fig.~\ref{ScatGeom}(a).

Measurements were carried out at $T = 50 \pm 10$~mK, with a range of applied magnetic fields,  $0.4$~T $\leq \mu_0 H \leq 2.6$~T, by using a dilution refrigerator insert in a horizontal-field cryomagnet. A motorized $\Omega$ stage rotated the dilution refrigerator around the vertical axis (crystalline $b$-axis) within the magnet, allowing measurements as the magnetic field was rotated within the crystalline $ac$-plane. The direction of the magnetic field was close to parallel to the incident neutron beam. A schematic of the experimental configuration is shown in Fig.~\ref{ScatGeom}(b).

The VL was prepared by first rotating the sample to the desired orientation ($\Omega$) and then changing $H$, followed by a damped small-amplitude field modulation with initial amplitude 20~mT. This method is known to produce a well-ordered VL in {\KFA} while remaining at the measurements (base) temperature, and eliminates the need for a time consuming field-cooling procedure before each measurement. 

The experiment was carried out using the SANS-I instrument at the Swiss Spallation Neutron Source (SINQ), the Paul Scherrer Institut, Switzerland. The measurements used neutron wavelengths $\lamn = 0.8$~nm or 1.2~nm and a bandwidth $\Delta \lamn/\lamn = 10\%$. A position sensitive detector, placed 11-18~m from the sample, was used to collect the diffracted neutrons. In order to satisfy the Bragg condition for the VL, the sample and magnet were tilted about the horizontal axis perpendicular to the beam direction [angle $\varphi$ in Fig.~\ref{ScatGeom}(b)]. Background measurements obtained in zero field were subtracted from the data.

\section{Results}
\begin{figure*}
  \includegraphics{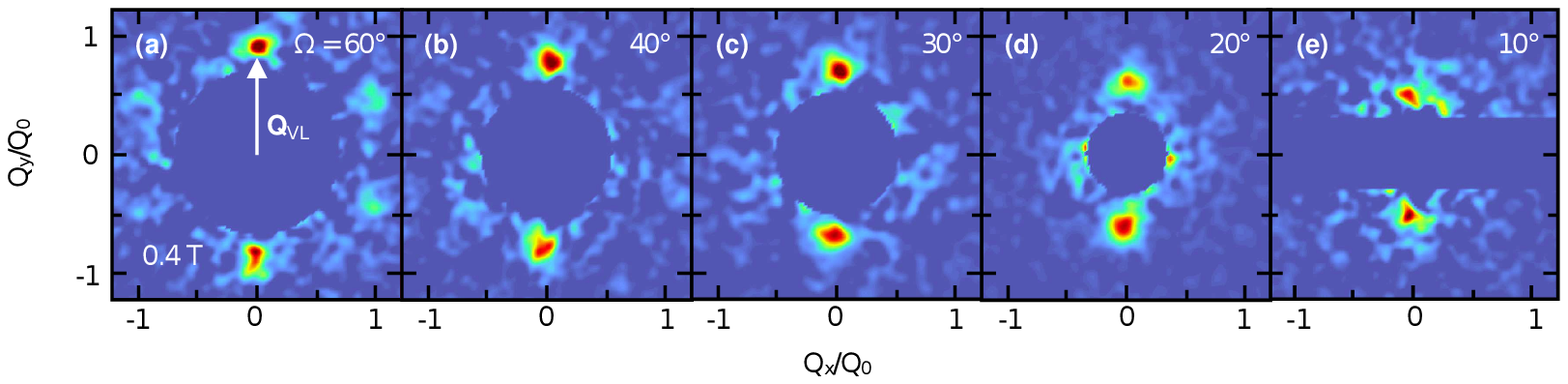}
  \includegraphics{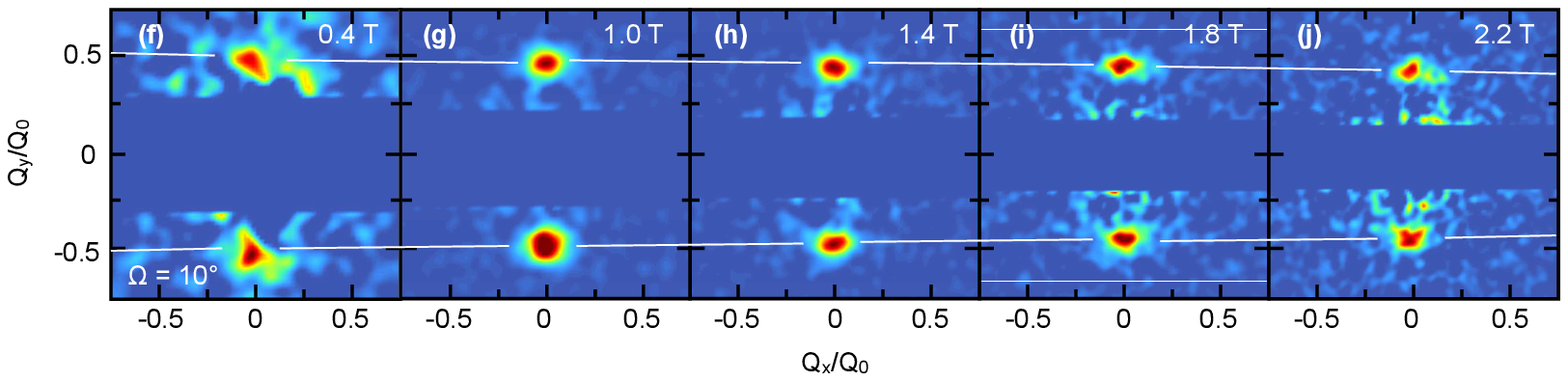}
  \caption{(Color online)
    Vortex lattice Bragg reflections. Positions in reciprocal space (horizontal and vertical axes) are normalized by the scattering vector for an isotropic triangular VL, $Q_0 = 1.075 \times 2\pi \sqrt{\mu_0 H/\fq}$.
    The diffraction patterns are the sum of measurements as the cryomagnet and sample mosaic are tilted about the horizontal axis perpendicular to the beam direction,
    which primarily satisfy the Bragg condition for reflections on the vertical axis ($Q_x =0$);
    however the remaining four first-order VL reflections with $Q_x/Q_0 \approx \pm 1$ and $Q_y/Q_0 \approx \pm 0.5$ are visible in (a).
    Panels (a) to (e): Measurements performed as an applied field of $0.4$~T was rotated towards the crystalline basal plane.    
    Panels (f) to (j): Measurements performed as a function of magnetic field at a fixed $\Omega = 10^{\circ}$.
    Panel (f) is identical to panel (e).
    The VL anisotropy increases with increasing field as indicated by the white line.
    The color scale is adjusted separately for all 9 panels to make the VL scattered intensity clearly visible.
    Imperfectly subtracted background scattering unrelated to the VL is masked off: In (a-d) in a circular region around $Q = 0$ and in (e-i) in a horizontal region around $Q_y = 0$
    (at small $\Omega$ the neutron beam is close enough to the basal plane to cause increased horizontal background scattering from crystal stacking faults and from the aluminum mounting plates).
   \label{DifPat}}
\end{figure*}

The diffraction patterns in Fig.~\ref{DifPat} show the VL Bragg peaks used to determine the superconducting anisotropy in {\KFA}.
Ideally, $\Gac$ is determined from measurements with the applied field parallel to the crystalline basal plane. In this configuration the VL Bragg peaks lie on an ellipse in reciprocal space with a major-to-minor axis ratio given by $\Gac$. However, due to the platelike crystal morphology, as well as generally weak scattering for this field orientation, such measurements are not possible. Instead we determine the VL anisotropy ($\GVL$) defined in Fig.~\ref{ScatGeom}(c) with the field applied at several angles with respect to the basal plane, Fig.~\ref{DifPat}(a-e). Extrapolation of the results makes it possible to obtain $\Gac = \GVL(\Omega = 0)$. Furthermore, as each vortex carries a single quantum of magnetic flux $\fq = h/2e = 2068$~T$\,$nm$^2$, the reciprocal space area of the ellipse is given by $\pi Q_0^2$ with $Q_0 = 2\pi (2 \mu_0 H/\sqrt{3}\fq)^{1/2}$. Thus, to determine $\GVL$ it is sufficient to measure VL Bragg reflections along one axis of the ellipse. For reasons to be discussed later we have used the reflections on minor axis, Fig.~\ref{ScatGeom}(c).

The magnitude of the minor axis scattering vector $\QVLvec$ [Fig.~\ref{DifPat}(a)] is equal to $Q_0$ for an isotropic VL, but decreases with increasing anisotropy such that $\GVL = (Q_0/\QVL)^2$. This is directly evident from panels (a) to (e) in Fig.~\ref{DifPat}, where the VL Bragg reflections move closer to $Q = 0$ as a constant applied field of $0.4$~T is rotated toward the basal plane, from $\Omega = 60^{\circ}$ to $10^{\circ}$. In addition to the $\Omega$-dependence, a field dependence of the VL anisotropy was found, shown in Fig.~\ref{DifPat}(f-j). In this case it necessary to separate the effect of a changing superconducting anisotropy from the increasing vortex density due to the change in the applied field. To achieve this, the axes in Fig.~\ref{DifPat} have all been normalized by $Q_0$. Plotted in this fashion it is apparent that $\GVL$ increases with increasing field, as indicated by the guide to the eye (white line) in panels (f) to (j).

\begin{figure}
  \includegraphics{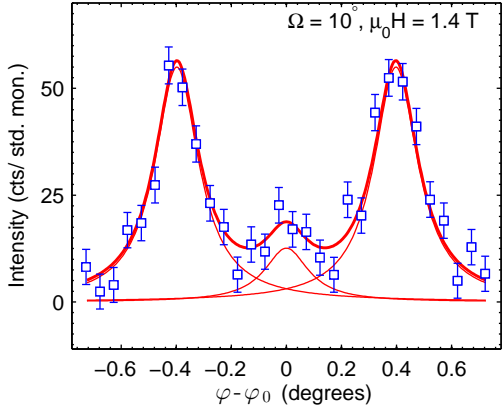}
  \caption{(Color online)
    Vortex lattice rocking curve, showing the scattered intensity as a function of sample tilt angle ($\varphi$) relative to the rocking curve center, $\varphi_0 = 0.28^{\circ}$.
    Three clear maxima are observed: A smaller center peak due to the longitudinal field modulation (non-spin flip) and a pair of stronger, Zeeman-split peaks due to the transverse field modulation (spin flip).
    The rocking curve is fitted by three Lorentzians, each with a width of $0.19^{\circ}$ FWHM,
    and with the two Zeeman split peaks located symmetrically around the center at $\varphi - \varphi_0 = \pm 0.40^{\circ}$.
    The neutron wavelength was $\lamn = 0.8$~nm.
    \label{RC}}
\end{figure}

Figure~\ref{RC} shows the evolution of the scattered intensity as a function of the tilt angle $\varphi$ (rocking curve) for a single VL Bragg peak at $1.4$~T and $\Omega = 10^{\circ}$.
The rocking curve is unusual in that it shows two large and one small peak, instead of a single maximum.
The magnitude of the VL scattered intensity is determined by the amplitude of the field modulation and is proportional to $|\bm{h}|^2$, where the form factor, $\bm{h}(\bm{q})$, is the Fourier transform of the magnetic induction, $\bm{B}(\bm{r})$.\cite{Eskildsen:2011jp}
In typical VL SANS experiments the form factor is due solely to the modulation of the longitudinal component of $\bm{B}(\bm{r})$ in the plane perpendicular to the applied field direction, denoted $h_z$ in Fig.~\ref{ScatGeom}(c).
In highly anisotropic superconductors, however, there is a strong preference for the vortex screening currents to run within the basal $ab$-plane. As the angle $\Omega$ between the applied field and the basal plane becomes small, the associated transverse field modulation, labeled $h_x$ in Fig.~\ref{ScatGeom}(c), becomes dominant.\cite{Thiemann:1989uw,Amano:2014ws}
The latter leads to spin flip (SF) scattering of the neutrons, and a Zeeman splitting of the VL rocking curves gives rise to the two large peaks at $\varphi = \varphi_0 \pm \Delta \varphi$.\cite{Rastovski:2013hh}
Less intense, non-spin flip (NSF) scattering due to $h_z$ is still visible at $\varphi = \varphi_0$.
Similar effects have previously been observed in yttrium barium copper oxide (YBCO)\cite{Kealey:2001fi} and in a more extreme form in strontium ruthenate.\cite{Rastovski:2013hh}

We now return to the determination of the intrinsic superconducting anisotropy of \KFA. As discussed above, the VL scattering in {\KFA} is dominated by the contribution from the transverse field modulation when the applied field is close to the basal plane. Furthermore, VL Bragg peaks that are not on the vertical axis have scattering vectors essentially parallel to $h_x$ [open circles in Fig.~\ref{ScatGeom}(c)], and are effectively unmeasurable as only components of the magnetization perpendicular to $\QVLvec$ will give rise to scattering.\cite{Squires:1978vi} For this reason, as well as the strong horizontal background scattering for small $\Omega$, the vertical VL Bragg peaks were used to determine the anisotropy.

\begin{figure}
  \includegraphics{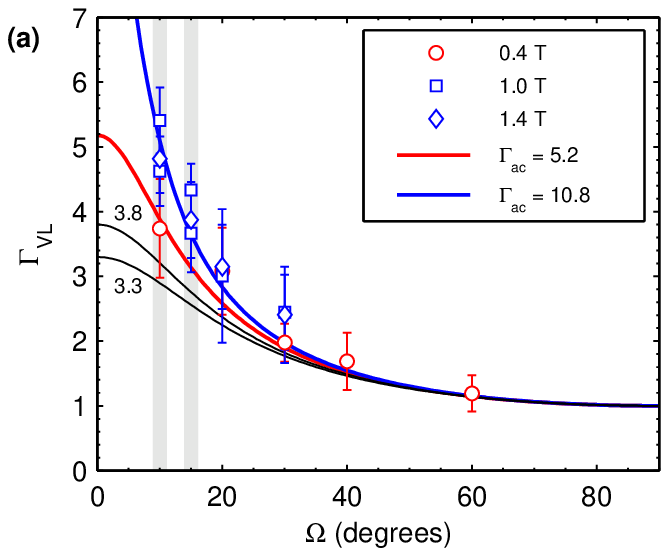}
  \includegraphics{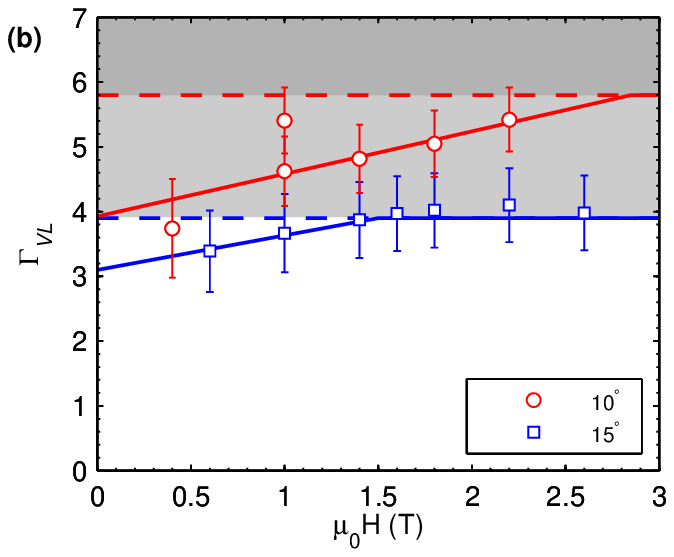}
  \caption{(Color online)
   Vortex lattice anisotropy as a function of the direction and magnitude of the applied magnetic field.
   Panel (a): $\GVL$ versus field rotation angle for different values of the applied magnetic field.
   The full red ($\Gac = 5.2$) and blue ($10.8$) curves are best fits to Eq.~(\ref{Geq}) for the $0.4$~T and the combined $1.0$ and $1.4$~T data respectively.
   The black lines correspond to $\GHcii = 3.3$\cite{Burger:2013tg} and $3.8$ obtained previously from low field SANS measurements.\cite{KawanoFurukawa:2013cs}
   The vertical grey bars represent the $\Omega$'s for which $\GVL$ is plotted below.
   Panel (b): Vortex lattice anisotropy as a function of the applied field for the two smallest values of the rotation angle.
   Full lines show fits to the data below the maximum possible value of $\GVL = (\sin \Omega)^{-1}$.
   \label{Anisotropy}}
\end{figure}

Figure~\ref{Anisotropy}(a) shows the VL anisotropy, $\GVL = (Q_0/\QVL)^2$, as a function of field orientation for applied fields of $0.4$, $1.0$, and $1.4$~T. 
In addition to determining the magnitude if the VL scattering vector from the peak position on the detector, as shown in Fig.~\ref{DifPat}(a), independent values were obtained from both the center of the rocking curve and the Zeeman splitting.
For the rocking curve center, obtained from the midpoint between the two Zeeman split peaks,
we have the usual Bragg's law in the small angle limit $\QVL = 2k_0 \varphi_0$, where $k_0 = 2\pi/\lamn$.
For the Zeeman splitting one obtains $\QVL = (2 k_0/\Delta \varphi) (\Delta \varepsilon/\varepsilon_0)$ with $\Delta \varepsilon = \gamma \mu_N B$ and $\varepsilon_0 = \hbar^2 k_0^2/2m_n$ where
$\gamma = 1.913$ is the neutron gyromagnetic ratio, $\mu_N = e\hbar/2m_n = 31.5$~neV/T is the nuclear magneton and $m_n$ is the neutron mass.\cite{Rastovski:2013hh}
Within experimental error the three methods agree, and the average value is shown in Fig.~\ref{Anisotropy}.
The data are fitted to the expression
\begin{equation}
  \GVL = \frac{\Gac}{\sqrt{\cos^2 \Omega + ( \Gac \, \sin \Omega )^2}}
  \label{Geq}
\end{equation}
obtained for a 3-dimensional superconductor with uniaxial anisotropy.\cite{Campbell:1988vf}
Although {\KFA} is a layered material the coherence length along the $c$-axis, $\xi_c = 2.45$~nm,\cite{Terashima:2009bi}
is still several times greater than the Fe-As interlayer spacing of $0.69$~nm,\cite{RefWorks:778} and we expect Eq.~(\ref{Geq}) to be applicable.
The fits yield values of the superconducting anisotropy $\Gac = 5.2 \pm 1.8$ at $0.4$~T, and $10.8 _{- 4.7}^{+21.9}$ for the combined high field data at $1.0$ and $1.4$~T (the large slope of the fitted high field curve gives rise to an asymmetric error).

To further elucidate a possible field dependence of the superconducting anisotropy, $\GVL$ is plotted versus $\mu_0 H$ in Fig.~\ref{Anisotropy}(b) for the two field orientations closest to the basal plane (smallest $\Omega$).
In both cases the VL anisotropy increases with increasing field, but the effect is most prominent for $\Omega = 10^{\circ}$.
For $\Omega = 15^{\circ}$ and fields above $1.4$~T, $\GVL$ saturates at $(\sin \Omega)^{-1}$, corresponding to an infinite $\Gac$ in Eq.~(\ref{Geq}).
When $\Omega \geq 20^{\circ}$ the dependence of the VL anisotropy on $\Gac$ become too small to measure accurately by SANS,
seen by the vanishing separation between the curves in Fig.~\ref{Anisotropy}(a) at larger $\Omega$.

\begin{figure}
  \includegraphics{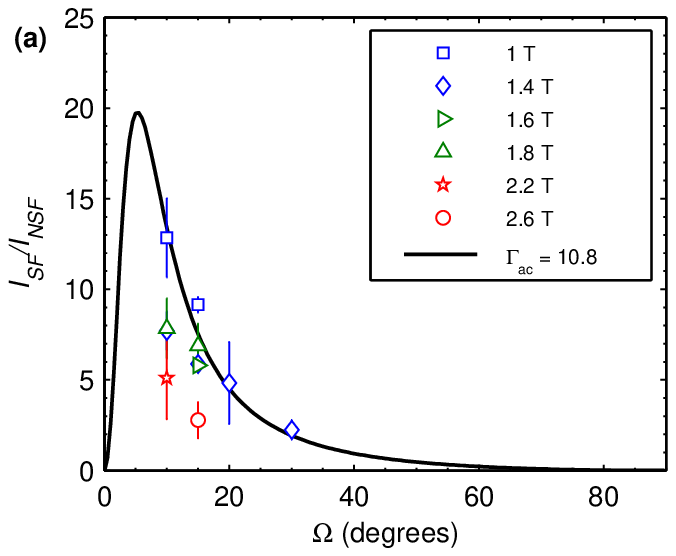}
  \includegraphics{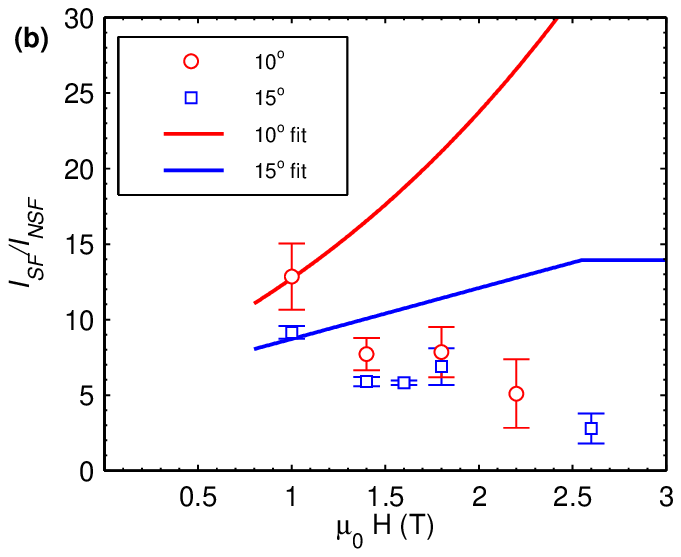}
  \caption{(Color online)
   Measured intensity ratio of spin flip to non-spin flip scattering as a function of the direction (a) and magnitude (b) of the applied magnetic field.
   Since the two SF peaks in the rocking curve (Fig.~\ref{RC}) each correspond to a different incident neutron spin,
   the ratio is the sum of their integrated intensities divided by the integrated intensity for the NSF peak which contains contribution from both spin directions.
   The line in panel (a) is calculated using Eq.~(\ref{SFNSFratio}) and the fitted high field value of $\Gac$ from Fig.~\ref{Anisotropy}(a).
   Similarly, the lines in panel (b) show the expected intensity ratio from the London model, based on the fitted field dependence of $\GVL$ from Fig.~\ref{Anisotropy}(b).
   \label{Intensity}}
\end{figure}

The superconducting anisotropy is also reflected in the the form factors $|h_x|$ and $|h_z|$. In principle these may be determined separately, using the integrated intensities of the respective peaks in the rocking curve, Fig.~\ref{RC}. When normalized to the incident neutron flux, this yields the VL reflectivity which can be related directly to form factors.\cite{Eskildsen:2011jp} In the present case, however, this is not feasible because the effective sample area and thickness depend on $\Omega$, and, more importantly, part of the neutron beam may pass between the mounting plates for fields close to the basal plane. Still, for a given $\Omega$, the ratio of the SF and NSF scattered intensity may be measured accurately by adding the integrated intensity of the Zeeman split peaks in the rocking curve and dividing the sum by the intensity of the peak at $\varphi = \varphi_0$, Fig.~\ref{Intensity}.

The intensity ratio may be compared to a London model calculation:\cite{Thiemann:1989uw}
\begin{eqnarray}
  \left(\frac{\ISF}{\INSF}\right)_{\text{London}}
    & = & \frac{|\hxL|^2}{|\hzL|^2} \nonumber \\
    & = & \left[ \frac{(1 - \Gac^2) \sin \Omega \cos \Omega}{\cos^2 \Omega + \Gac^2 \sin^2 \Omega} \right]^2.
  \label{SFNSFratio}
\end{eqnarray}
Since the London model assumes an unphysical $\delta$-function vortex core, the expressions for  $|h_x|$ and $|h_z|$ are typically multiplied by a so-called core correction factor, $\exp[-c \, Q^2 \, \xi^2]$, where $c$ is a constant of order unity and $\xi$ is the coherence length.\cite{Eskildsen:2011jp}
For the intensity ratio in Eq.~(\ref{SFNSFratio}) the core correction for the two form factors divide out and as a result the only field dependence will be through $\Gac$.
This is confirmed by more sophisticated numerical solutions to the Eilenberger equations that only find minor corrections to the London intensity ratio.\cite{Amano:2014ws}

Inserting the fitted high field value of $\Gac$ from Fig.~\ref{Anisotropy}(a) in Eq.~(\ref{SFNSFratio}) yields a calculated intensity ratio that accurately describes the 1~T measurements with no adjustable parameters, as shown in Fig.~\ref{Intensity}(a). However, with increasing field the measured intensity ratio falls further and further below the calculated London model curve. 
The discrepancy between the measured and calculated values is even more evident in the field dependence of  $\ISF/\INSF$, Fig.~\ref{Intensity}(b).
The measured intensity ratio decreases with increasing field, in stark contrast to the calculated $\ISF/\INSF$ based on the field dependence of $\GVL$ from Fig.~\ref{Anisotropy}(b).
The latter depends only on the VL anisotropy, and therefore $\Gac$, and is expected to grow with increasing field, potentially reaching a constant value when or if $\GVL$ saturates. 
In order to explain the measured $\ISF/\INSF$ is it clearly necessary to go beyond a simple London model, and we will return to this issue later.

Finally, as an aside, we note that  the transverse magnetization will cause a rotation of the magnetic induction, $\bm{B}$, relative to the direction of the applied field, $\bm{H}$, and hence a difference between the nominal and actual values of $\Omega$. The transverse magnetization is related to the transverse VL form factor, $h_x$.\cite{RefWorks:786,Thiemann:1989uw}
Previous {\KFA} SANS studies with $\bm{H} \parallel \bm{c}$ found a longitudinal VL form factor $h_z \leq 1$~mT,\cite{KawanoFurukawa:2011hw} which provides an upper limit on $h_x$ for fields close to the basal plane.\cite{Thiemann:1989uw,Rastovski:2013hh}
From this we find that the ``misalignment'' between the nominal and actual value of $\Omega$ is less than $1^{\circ}$.
This order of magnitude is consistent with a calculation that, in addition to the magnetization, also includes the demagnetization effects due to the platelike crystal morphology, and which was previously used successfully to model VL SANS data obtained for {\MgB}.\cite{RefWorks:12}
The horizontal width of the data points in Figs.~\ref{Anisotropy}(a) and \ref{Intensity}(a) is approximately $\pm 1^{\circ}$, comparable to the maximum possible error on the actual value of $\Omega$. This uncertainty does not affect our analysis of the SANS data in any significant way.

\section{Discussion}
In superconducting KFe$_2$As$_2$, $H_{c2}$ (in Tesla) parallel to the basal plane is larger than $T_c$ (in Kelvin) suggesting Pauli limiting.
Measurements of the upper critical field anisotropy at low temperature, comparable to those used in our SANS experiments, consistently yield a value $\GHcii = 3.3$.\cite{Burger:2013tg,Zocco:2013gz}
This should be contrasted to the extrapolation of the measured VL anisotropy to $\Omega = 0$, which provides a direct measure of the intrinsic superconducting anisotropy, $\Gac$.
As shown in Fig.~\ref{Anisotropy}, the latter increases with field, and already at $0.4$~T it exceeds $\GHcii$.
In single band superconductors where the upper critical field is orbitally limited for all field directions one expects $\Gac = \GHcii$.
Our data show that $\Hcii$, for fields within and close to the basal plane, is suppressed below the orbital limit. This is consistent with zero temperature estimates of the orbital upper critical field obtained from an extrapolation of the slope $(\partial \Hcii/\partial T)_{T = \Tc}$, suggesting strong Pauli limiting for fields within the basal plane.\cite{Burger:2013tg,Zocco:2013gz}
In addition, $\Hcii$ becomes a first order transition below $1.5$~K for $\bm{H} \perp \bm{c}$,\cite{Zocco:2013gz} which is also indicative of strong Pauli paramagnetic effects (PPEs).
At higher temperature, where PPEs effects decrease,\cite{RefWorks:756} the upper critical field anisotropy is expected to increase towards $\Gac$.
This is confirmed experimentally, with $\GHcii$ reaching $\sim 7$ at the critical temperature.\cite{Zocco:2013gz}
This higher value also agrees well with the reported ratios of $(\partial \Hcii/\partial T)$ close to $\Tc$ for fields within and perpendicular to the basal plane, which are in the range $5.4 - 6.2$.\cite{Liu:2013hy,Burger:2013tg}
We note that a $T$-dependence of $\GHcii$ may also be due to multiband superconductivity,\cite{Kogan:2012hf} and we will return to a discussion of multiband effects later.
However, below we will first provide additional support for PPEs in {\KFA}.

\begin{figure} 
  \includegraphics{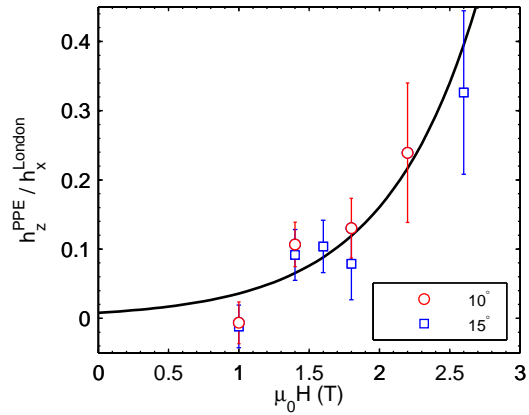}
  \caption{(Color online)
   Fractional contribution of Pauli paramagnetic effects to the form factor $h_z$ parallel to the vortices obtained using Eq.~(\ref{PPEdef}) and the measured and calculated values of $\ISF/\INSF$ from Fig.~\ref{Intensity}(b).
   The line is a simple exponential fit to the data.
   \label{PPE}}
\end{figure}

While suppression of the upper critical field is a strong indication of Pauli limiting, clear evidence of the effects of Pauli paramagnetism on the unpaired quasiparticles in the vortex cores is directly evident from the measured ratio of spin flip to non-spin flip scattering in Fig.~\ref{Intensity}. As already discussed, the $\Omega$ dependence, which at 1~T agrees well with the simple London model expression in Eq.~(\ref{SFNSFratio}),  deviates substantially at higher fields. Likewise, $\ISF/\INSF$ decreases with increasing $H$ in contrast to the expected constant or increasing behavior.
Previously, it has been shown that strong PPEs will lead to a substantial polarization of the unpaired quasiparticle spins in the vortex cores.\cite{RefWorks:6,RefWorks:756}
The periodicity of this spin polarization is inherently commensurate with the VL field modulation due to the superconducting screening currents, and gives rise to an additional contribution to the longitudinal form factor, $h_z = h_z^{\text{London}} + h_z^{\text{PPE}}$.
As a result we expect 
\begin{equation*}
  \left( \frac{\ISF}{\INSF} \right)_{\text{meas}}
    = \frac{|\hxL|^2}{|\hzL + \hzPPE |^2},
\end{equation*}
which allows us to separate the contribution of the PPE by the expression
\begin{equation}
   \frac{|\hzPPE|}{|\hxL|}
     =  \left( \frac{\ISF}{\INSF} \right)_{\text{meas}}^{-1/2} - \left( \frac{\ISF}{\INSF} \right)_{\text{London}}^{-1/2}.
  \label{PPEdef}
\end{equation}
Figure~\ref{PPE} shows the PPE contribution as a function of field obtained in this fashion.
This increases with $H$ over the entire measured field range, and is well fitted by a simple exponential function. Below 1~T the Zeeman splitting of the SF peaks is too small to allow a reliable fitting of the 3 peaks in the rocking curve. Due to the decreasing VL scattering with increasing field we are not able to image the VL above $2.6$~T. However, at higher fields the PPEs are expected to saturate and eventually disappear abruptly at the first order $\mu_0 \Hcii = 5$~T.\cite{RefWorks:1,Amano:2015vd}

The ability to measure the PPEs as a function of both field and temperature deep within the superconducting phase is a great advantage of the SANS technique, and as far as we know unique.
It provides a detailed, quantitative measure of the strength of PPE that could, in principle, be compared directly to numerical calculations. We note that our analysis ignores the weak magnetic anisotropy of {\KFA}.\cite{Hardy:2013fe} While it would be straightforward to include this in the analysis, the results would not change appreciably since $\hxPPE/\hzPPE \alt (\chi_{ab}/\chi_c - 1) \sin\Omega \ll 1$.
The resulting corrections are $\leq 4\%$, and thus insignificant compared to the precision with which we can determine the PPE contribution to the VL form factors.
Finally, we note that in the present case $\hzPPE$ only has a significant magnitude for fields close to the basal plane where Pauli limiting of $\Hcii$ is observed.
This is in contrast to CeCoIn$_5$ where equal PPEs were observed for both in- and out-of-plane fields.\cite{Das:2012fb}

The field dependence of $\GVL$ in Fig.~\ref{Anisotropy}(b), and by extension $\Gac$, is a strong indication of multiband superconductivity. The intrinsic superconducting anisotropy arises from the Fermi surface sheets that carry the superconducting Cooper pairs, and in a single band superconductor it is determined by the ratio of Fermi velocities $\Gac = v_{ab}/v_c$. For a multiband superconductor an intermediate value is expected, lying in the range spanned by the individual supercurrent carrying sheets.
If the energy gaps for the different bands differ in amplitude so will their sensitivity to the applied magnetic field, and increasing $H$ will change the relative contribution to the superconductivity.
In cases where the bands have different Fermi velocity ratios this will lead to a field dependence of the $\Gac$, as first demonstrated in the case of {\MgB}.\cite{RefWorks:19}

Our conclusion of multiband superconductivity is in good agreement with other studies of {\KFA}.
De Haas-van Alphen measurements show three concentric Fermi surface hole cylinders ($\alpha$, $\beta$, $\zeta$) around the Brillouin Gamma-point
and a propeller-like sheet ($\varepsilon$) around the M-point.\cite{Terashima:2010iu,Terashima:2013vv}
Furthermore, they showed a smaller anisotropy for the $\zeta$ sheet and a larger anisotropy for the $\alpha$, $\beta$ and $\varepsilon$ sheets.
A wide range of band-specific superconducting gaps 
in {\KFA} were derived from combined specific heat and magnetization studies,\cite{Hardy:2014te}
and are in excellent agreement with previous SANS VL measurements.\cite{KawanoFurukawa:2011hw}
Superconducting gap nodes were also observed by ARPES,
although with gap amplitudes much greater than those obtained by other techniques and expected from Bardeen-Cooper-Schrieffer theory.\cite{Okazaki:2012bf}
Importantly, superconductivity is suppressed by fields $H \ll  \Hcii$ on the bands with the smaller energy gaps.\cite{Hardy:2014te}
This explains why we observe the change in $\Gac$ for fields much below $\Hcii$, regardless of whether one consider the Pauli (5~T) or orbitally limited ($9 - 15$~T) upper critical field.
Once again, this situation is similar to what was previously observed in {\MgB}.\cite{Eskildsen:2002ih,RefWorks:19,Tewordt:2003fl}

\section{Conclusion}
In conclusion, we have used SANS to study the anisotropy of the superconducting state as well as Pauli paramagnetic effects in {\KFA}, extending previous studies to higher applied magnetic fields and directions closer to the basal plane. This has allowed a more precise determination of the superconducting anisotropy $\Gac$ which was found to be strongly field dependent, providing strong support for multiband superconductivity. Moreover, $\Gac$ exceeds the upper critical field anisotropy, indicating Pauli limiting in {\KFA} for field applied within the basal plane. Finally, we were able to directly quantify the contribution to the field modulation in the mixed state due to PPEs by separately measuring both the non-spin flip and spin flip VL scattered intensity. Our studies represents the most comprehensive SANS study to date of multiband superconductivity and the effects of Pauli paramagnetism in any superconductor.

\section*{Acknowledgements}
We acknowledge valuable discussions with K. Machida.
This work was supported by
the U.S. Department of Energy, Office of Basic Energy Sciences, under Award No.~DE-FG02-10ER46783 (SJK, MRE);
the Institute for Solid State Physics, University of Tokyo and originally approved (proposal no. 14573) for JRR-3, Japan Atomic Energy Agency (HKF);
the European Commission under the 7th Framework Programme through the 'Research Infrastructures' action of the 'Capacities' Programme, NMI3-II Grant number 283883 and
the UK EPSRC under Grant number EP/J016977 (EJ, RR and EMF);
the Japan Society for the Promotion of Science, under Grant-in- Aid for Scientific Research B No.~24340090.
This work is based on experiments performed at the Swiss spallation neutron source SINQ, Paul Scherrer Institute, Villigen, Switzerland.

\bibliography{KFAanisotropy}

\end{document}